\documentstyle[12pt]{article}
\newcommand{\es}{\\[2mm]}

\newcommand{\journal}[4]{{\em #1~}#2\,(19#3)\,#4;}

\newcommand{\hpa}{\journal {Helv. Phys. Acta}}

\newcommand{\ijmp}{\journal {Int. J. Mod. Phys.}}

\newcommand{\pr}{\journal {Phys. Rev.}}

\newcommand{\jmp}{\journal {J. Math. Phys.}}

\newcommand{\cmp}{\journal {Comm. Math. Phys.}}
\newcommand{\cqg}{\journal {Class. Quantum Grav.}}

\newcommand{\np}{\journal {Nucl. Phys.}}
\newcommand{\pl}{\journal {Phys. Lett.}}

\newcommand{\nc}{\journal {Nuovo Cim.}}

\newcommand{\annp}{\journal {Ann. Phys. (N.Y.)}}

\makeatletter
\@addtoreset{equation}{section}
\makeatother

\def\LP{\displaystyle{\Biggl(}}
 \def\wti{\widetilde}
\def\RP{\displaystyle{\Biggr)}}
\newcommand{\lp}{\left(}\newcommand{\rp}{\right)}
\newcommand{\lc}{\left[}\newcommand{\rc}{\right]}

\renewcommand{\d}{\delta}

\newcommand{\g}{\gamma}

\newcommand{\m}{\mu}

\renewcommand{\o}{\omega}

\newcommand{\dT}{\wti d}
\newcommand{\DT}{\wti {\cal D}}
\newcommand{\BT}{\wti {\cal B}}
\renewcommand{\AA}{{\cal A}}
\newcommand{\BB}{{\cal B}}

\newcommand{\FF}{{\cal F}}
\newcommand{\GG}{{\cal G}}

\newcommand{\LL}{{\cal L}}
\newcommand{\MM}{{\cal M}}
\newcommand{\NN}{{\cal N}}

\newcommand{\QQ}{{\cal Q}}

\newcommand{\VV}{{\cal V}}
\newcommand{\WW}{{\cal W}}

\newcommand{\complex}{{\kern .1em {\raise .47ex
\hbox {$\scriptscriptstyle |$}}
    \kern -.4em {\rm C}}}
\newcommand{\real}{{{\rm I} \kern -.19em {\rm R}}}
\newcommand{\rational}{{\kern .1em {\raise .47ex
\hbox{$\scripscriptstyle |$}}
    \kern -.35em {\rm Q}}}
\renewcommand{\natural}{{\vrule height 1.6ex width
.05em depth 0ex \kern -.35em {\rm N}}}
\newcommand{\dint}{\displaystyle{\int}}

\newcommand{\tr}{{\rm {Tr} \,}}

\newcommand{\pa}{\partial}

\newcommand{\dfrac}[2]{{\displaystyle{\frac{#1}{#2}}}}
\newcommand{\dsum}[2]{\displaystyle{\sum_{#1}^{#2}}}

\newcommand{\ie}{{{\em i.e.}\ }}

\newcommand{\sla}{\raise.15ex\hbox{$/$}\kern -.57em}

\newcommand{\twiddle}{\lower.9ex\rlap{$\kern -.1em\scriptstyle\sim$}}

\newcommand{\vf}{{\varphi}}
\newcommand{\bpart}{{\bar \partial}}

\newcommand{\equ}[1]{(\ref{#1})}

\newcommand{\eq}{\begin{equation}}
\newcommand{\eqn}[1]{\label{#1}\end{equation}}
\newcommand{\eea}{\end{eqnarray}}
\newcommand{\eqa}{\begin{eqnarray}}
\newcommand{\eqan}[1]{\label{#1}\end{eqnarray}}
\newcommand{\ba}{\begin{array}}
\newcommand{\ea}{\end{array}}
\newcommand{\eqac}{\begin{equation}\begin{array}{rcl}}
\newcommand{\eqacn}[1]{\end{array}\label{#1}\end{equation}}

\newcommand{\at}{{\~a}}

\newcommand{\ooo}{{\'o}}

\newcommand{\iii}{{\'\i}}

\newcommand{\ott}{{\wti \o}}

\begin{document}


{\ }

\vspace{3cm}
\centerline{\LARGE BRS Cohomology of Zero Curvature Systems}\vspace{2mm}
\centerline{\LARGE I. The Complete Ladder Case}

\vspace{1cm}

\centerline{\bf {\large M. Carvalho, L.C.Q. Vilar, C.A.G. Sasaki}}
\vspace{2mm}
\centerline{\it C.B.P.F}
\centerline{\it Centro Brasileiro de Pesquisas F{\iii}sicas,}
\centerline{\it Rua Xavier Sigaud 150, 22290-180 Urca}
\centerline{\it Rio de Janeiro, Brazil}
\vspace{3mm}
\centerline{ and }
\vspace{3mm}
\centerline{\bf {\large S.P. Sorella}}
\vspace{2mm}
\centerline{{\it UERJ}}
\centerline{{\it Departamento de F{\iii}sica Te{\ooo}rica}}
\centerline{{\it Instituto de F{\iii}sica, UERJ}}
\centerline{{\it Rua S{\at}o Francisco Xavier, 528}}
\centerline{{\it 20550-013, Rio de Janeiro, Brazil}}
\centerline{ and }
\centerline{\it C.B.P.F}
\centerline{\it Centro Brasileiro de Pesquisas F{\iii}sicas,}
\centerline{\it Rua Xavier Sigaud 150, 22290-180 Urca}
\centerline{\it Rio de Janeiro, Brazil}

\vspace{4mm}


\centerline{{\normalsize {\bf REF. CBPF-NF-062/95}} }

\vspace{4mm}
\vspace{10mm}

\centerline{\Large{\bf Abstract}}\vspace{2mm}
\noindent
We present here the zero curvature formulation for a wide class of field
theory models. This formalism, which relies on the existence of an operator
$\d$ which decomposes the exterior space-time derivative as a BRS
commutator, turns out to be particularly useful in order to solve
the Wess-Zumino consistency condition.
The examples of the topological theories
and of the $B$-$C$ string ghost system are considered in detail.
\setcounter{page}{0}
\thispagestyle{empty}

\vfill
\pagebreak
\section{Introduction}

Nowadays it is an established fact that the search of the possible
anomalies and of the counterterms which arise at the quantum level in
local field theories can be done in a purely algebraic way\footnote{For a
recent account on the so called {\it Algebraic Renormalization}
see~\cite{book}.} by
identifying the cohomology classes of the nilpotent BRS operator $b$ in the
space of the integrated local polynomials in the fields and their derivatives.
This means that one has to look at the nontrivial solutions of the equation
\eq
             b \int \o_D^G = 0   \ ,
\eqn{brs-cons-cond}
$\o_D^G$ denoting a local polynomial in the fields of ghost number $G$ and
form degree $D$, $D$ being the dimension of the space-time.
The cases $G=0,1$ correspond respectively to counterterms and anomalies.

The BRS consistency condition \equ{brs-cons-cond}, when translated at the
nonintegrated level,
yields a system of equations usually called descent equations (see~\cite{book}
and refs. therein)
\eq\ba{l}
    b\, \o^G_{D{\ }{\ }{\ }}         + d\, \o^{G+1}_{D-1}  = 0 \ , \es
    b\, \o^{G+1}_{D-1} + d\,\o^{G+2}_{D-2}  = 0 \ , \es
     \qquad  ...........                             \es
     \qquad  ...........                              \es
    b\, \o^{G+D-1}_{1} + d\, \o^{G+D}_{0}    = 0 \ , \es
    b\, \o^{G+D}_{0}                       = 0 \ ,
\ea\eqn{desc-eq}
$d=dx^{\m} \pa_{\m}$ being the exterior space-time derivative and
$ \o^{G+D-j}_{j} ( \, 0 \leq j \leq D )$ local polynomials of ghost
number $(G+D-j)$ and form degree $j$. The operators $b$ and $d$ obey the
algebraic relations
\eq
    b^2\, = \, d^2\, = \, b\,d + d\,b \, = 0   \ .
\eqn{b-d-algebra}
It should be remarked that at the nonintegrated level one looses the
property of making integration by parts. This implies that the fields
and their derivatives have to be considered as independent variables.

The problem of solving the descent equations (\ref{desc-eq}) is a problem
of cohomology of $b$ modulo $d$, the corresponding cohomology classes being
given by solutions of \equ{desc-eq} which are not of the type
\eq\ba{l}
     \o^{G+D-m}_m  =  b\, \hat \o^{G+D-m-1}_m + d\,\hat \o^{G+D-m}_{m-1}
 \ , \qquad 1\leq m\leq D \, \es
      \o^{G+D}_{0} = b\,\hat \o^{G+D-1}_0 \ ,
\ea\eqn{triv-sol}
with $\hat \o$'s local polynomials.

Recently a new method for finding nontrivial solutions of the tower
\equ{desc-eq} has been proposed by one of the authors~\cite{dec} and
successfully
applied to a large number of field models such as Yang-Mills
theories~\cite{tat}, gravity~\cite{dec-grav}\footnote{In the case of
gravitational theories,
the decomposition $d=-\lc b,\d \rc$ was in fact already
observed, see for instance refs.~\cite{band,bran}.} , topological field
theories~\cite{topol1,topol2,topol3},
string~\cite{corde} and superstring~\cite{scorde} theories, as well as
$W_3$-algebras~\cite{w3-dec}.
The method relies on the introduction of an operator $\d$ which allows
to decompose the exterior derivative as a BRS commutator,
\ie
\eq
              d=-\lc b,\d \rc \ .
\eqn{delt-def}
It is easily proven that, once the decomposition \equ{delt-def}
has been found, repeated applications of the operator
$\d $ on the polynomial $\o^{G+D}_{0}$
which solves the last of the equations \equ{desc-eq} will give an explicit
nontrivial
solution for the higher cocycles $\o^{G+D-j}_{j}$.

One has to note that solving the last equation of the tower \equ{desc-eq}
is a problem of local cohomology instead of a modulo-$d$ one. Moreover, the
former can be systematically attacked by using several methods as, for
instance, the spectral sequences technique~\cite{spectr}.
It is also worthy to mention that the solutions of the descent equations
\equ{desc-eq} obtained via the decomposition \equ{delt-def} have been
proven to be equivalent to those provided by the transgression procedure
based on the so called {\it Russian Formula}~\cite{russian,russian1}.

The aim of the present work is twofold. First, to improve and extend
the results obtained in~\cite{dec} and, secondly, to discuss the deep relation
between the existence of the operator $\d$ entering the decomposition
\equ{delt-def} and the possibility of encoding all the relevant informations
(BRS transformations of the fields, BRS cohomology classes, solutions of the
descent equations) into a unique equation which takes the form of a
generalized zero curvature condition:
\eq
{\wti \FF} = \dT {\wti \AA}  -i  {\wti \AA}^2 =0 \ .
\eqn{zero-curv-cond}
The operator $\dT$ and the generalized gauge connection ${\wti \AA}$ in
eq.\equ{zero-curv-cond} turn out to be respectively the $\d$-transform
of the BRS operator $b$ and of the ghost field $c$ corresponding to
the Maurer-Cartan form of the underlying gauge algebra
\eq\ba{l}
\dT = e^{\d}\,b\,e^{-\d}=  b + d + {....}\ , \qquad {\dT}^{\,2}=0 \ , \es
{\wti \AA} = e^{\d}\,c = c + ....  \ .
\ea\eqn{dtilde-Atilde}
The main purpose of this work will be that of clarifying  the meaning of the
 generalized gauge connection ${\wti \AA}$ and of the dots $....$ appearing in
 equation \equ{dtilde-Atilde} with the help of several examples.

In particular, as we shall see, the zero-curvature condition
\equ{zero-curv-cond} immediately yields the cohomology classes of the
generalized nilpotent operator\footnote{The nilpotency of $\dT$
is a direct consequence of the zero curvature condition
\equ{zero-curv-cond}.} $\dT$~\cite{ggl}. The latters turn out to be
naturally related to the solutions of the descent equations \equ{desc-eq}.
In other words, once the zero curvature condition of the model under
consideration has been established, the problem of finding the anomalies
and the invariant actions becomes straightforward  to be solved.

For the sake of completeness and in order to present several
detailed models, the paper has been splitted in two parts, referred as
 part I and part II. This division corresponds to two different situations,
called respectively the {\it complete} and the {\it noncomplete} ladder case.
In the first case the components of the generalized gauge connection
${\wti \AA}$ form a ladder of fields which span all possible form degrees
compatible with the space-time dimension $D$. This means that,
ordering  the components of ${\wti \AA}$ according to their increasing
form degree
$p$, the allowed interval, \ie $0\leq p\leq D$, is fully covered.
Instead, in the noncomplete ladder case the maximum form degree reached by
the components of ${\wti \AA}$ is strictly lower than the space-time
dimension $D$, \ie $0\leq p < D$.

Examples of models belonging to the first case are, for instance, the
topological models of the Schwartz type such as the Chern-Simons and
the $BF$ models~\cite{topol-rep}, and the $B$-$C$ ghost system of the bosonic
string theory~\cite{gsw}.
On the other hand, the Yang-Mills type theories can be accommodated in the
noncomplete ladder case.

Even if many properties of the models covered by the complete ladder case
have already  been investigated~\cite{topol2,ncab}, their zero curvature
formulation still
represents a very elegant and interesting aspect.
Moreover, in the noncomplete case, the zero curvature condition
\equ{zero-curv-cond} requires the existence of a set of new operators
$(\GG^{1-k}_k , \,\, 2\leq k\leq D )$ which are in involution, i.e. the
operator $\GG^{1-k}_k$ is generated by the commutator between $\GG^{2-k}_{k-1}$
and the operator $\delta$ of \equ{delt-def}, according to the recursive
formula
\eq\ba{l}
\GG^{-1}_2 \, = \dfrac{1}{2}[\,{\d}\,,\, d\,] \ ,\, \\[3mm]
\GG^{1-k}_k = \dfrac{1}{k}\left [\,\d \,,\, \GG^{2-k}_{k-1}\,\right ]
\ , \qquad k > 2 \ .
\ea\eqn{invol-oper}

This  structure naturally reminds us to the recursive construction of the
Lax pair operators of the integrable systems~\cite{integr}.
This is a quite welcome and attractive feature which may signal a deeper
relation between the BRS cohomology techniques and the integrability.
Needless to say, the zero curvature condition represents in fact
one of the most important chapters of the integrable systems
(see also the recent works of ref.~\cite{zercurv}).

The part I of the paper ialgebraic set up is presented. In Sect. 3 we discuss
the geometrical
meaning of the zero curvature condition.
Sections 4 and 5 are devoted respectively to the computation of the BRS
cohomology and to characterize the solution of the descent equations.
Sect. 6 deals with the coupling with matter fields in the context of the
$BF$ models.
Without entering in details, let us briefly comment that, in analogy with
the gauge ladder ${\wti \AA}$, the matter fields can be introduced by
means of a second complete ladder $\BT$ constrained by the requirement
of being covariantly constant with respect to the gauge ladder, \ie
\eq
\DT \BT= \dT \BT - i [ {\wti \AA}, \BT ] = 0 \ .
\eqn{Bianchi}
As we shall see, condition \equ{Bianchi}
completely characterizes the BRS transformations of the various components
of $\BT$.

Finally, Sect. 7 contains a detailed discussion of the zero curvature
formulation of the $B$-$C$ string ghost system.

\section{The general set up}

In order to present the general algebraic set up let us begin by
fixing the notations.
We shall work in a space-time of dimension $D$ equipped with a
set of fields generically denoted by $\{\vf_q^p \}$, $q$ and $p$
being respectively the form degree and the ghost number.
The components $\vf_q^p$ will be treated as commuting or
anticommuting variables according to the fact that their total
degree, \ie the sum $(q+p)$, is even or odd. Otherwise stated,
the $\vf_q^p$'s are Lie-algebra valued, $\vf_q^p=(\vf_q^p)^aT^a$,
$T^a$ being the hermitian generators of a compact semisimple
Lie group $G$. Moreover, these fields are assumed to be collected into a
unique
generalized complete field ${\wti \AA}$ of total degree one, \ie
\eq
{\wti \AA} =\sum_{j=0}^D\,\vf^{1-j}_j=\vf^1_0 + \vf^0_1 + \vf^{-1}_2 + ...
+\vf^{1-D}_D \ .
\eqn{ladder}

The name complete, as already said in the introduction, is due to
the fact that the field content of the expansion \equ{ladder}
spans all possible
form degrees. In addition, eq.\equ{ladder} shows that the generalized field
${\wti \AA}$ contains a zero form with ghost number one $\vf^1_0$, and a
one-form with ghost number zero $\vf^0_1$. These fields will be
naturally identified with the Faddeev-Popov ghost field and with the
gauge connection of the familiar Yang-Mills gauge transformations.
Therefore ${\wti \AA}$ will be called the gauge ladder and the
components $\vf^1_0$ and $\vf^0_1$ will be denoted respectively by
$c$ and $A$, so that
\eq
{\wti \AA} = c + A + \vf^{-1}_2 + ...+ \vf^{1-D}_D \ .
\eqn{ladder-cA}

Finally, the functional space $\VV$ the BRS operator $b$ acts
upon is the space of the form-valued polynomials in the
fields $\vf^{1-j}_j$ and their differentials, \ie
\eq
\VV = \hbox{polynomials in}\,\,
\lp \vf^{1-j}_j,\,d\,\vf^{1-j}_j \,;\,\,\, 0\leq j \leq D \rp ,
\eqn{base}
$d$ being the exterior derivative defined\footnote{Observe
that $d\vf^{1-D}_D$ automatically vanishes, due to the dimension of the
space-time.} as
\eq
 d\eta_p = dx^{\mu} \partial_{\mu}\eta_p
\eqn{ddef}
for any $p$-form
\eq
\eta_p = \dfrac{1}{p!} \, \eta_{i_1.....i_p}dx^{i_1}.....\,dx^{i_p} \ ,
\eqn{eta-form-def}
where a wedge product has to be understood.

Let us also mention that, as  proven in~\cite{bdk,dvhtv}, the  space of
polynomials of forms turns out to be bigger in half to include the anomalies
and the Chern-Simons type actions which, as it is well known, can be naturally
written in terms of differential forms.

In order to obtain the BRS transformations of the fields
belonging to the gauge ladder \equ{ladder-cA} we introduce the
generalized operator of total degree one\footnote{The operators
$b$ and $d$ rby one unit.}
\eq
\dT = b + d,
\eqn{d-tilde2}
and we impose the zero curvature condition
\eq
\dT {\wti \AA} = i  {\wti \AA}^2 = \dfrac{i}{2}\lc {\wti \AA} \,,\,{\wti \AA}
\,\rc ,
\eqn{zero-curvature-cond}
where $\lc a , b \rc = ab-(-1)^{|a||b|}ba$ denotes the graded
commutator and $|a|$ is the total degree of $a$.

Developping equation \equ{zero-curvature-cond} in components and
identifying the terms with the same ghost number and form degree
we obtain the following transformations
\eq\ba{l}
  b  c= ic^2 \ , \es
  b  A= -d c + i [c,A] \ , \es
  b \vf^{1-j}_j =-d \vf^{2-j}_{j-1} + \dfrac{i}{2} \dsum{m=0}{j}
    \lc \vf^{1-m}_m\,,\,\vf^{1-j+m}_{j-m} \rc  \ ,
  \qquad   2 \leq j \leq D \ ,
\ea\eqn{brs-transform-cAladder}
which are easily checked to be nilpotent
\eq
 b^2=0 \ .
\eqn{nilpotency}
Notice that, as announced, the transformations of the first two
components of the ladder ${\wti \AA}$ are nothing but the familiar BRS
transformations of the Faddeev-Popov ghost and of the
Yang-Mills gauge connection.

\noindent
Let us introduce now the operator $\d$ defined by
(see also refs.~\cite{book,dec-grav,topol2})
\eq
{\wti \AA} = e^{\d}c,
\eqn{A-tilde2}
\ie
\eq\ba{l}
    \d \, \vf^{1-j}_j = (j+1)\vf^{-j}_{j+1}\,, \qquad 0\leq j \leq D-1 \,, \es
   \d \vf^{1-D}_D = 0 \ .
\ea\eqn{field-delta-action}
Its action extends on the differentials $(d\,\vf^{1-j}_j \ ,{\ } 0\leq j \leq
D )$ as
\eq\ba{l}
    \d \,d\, \vf^{1-j}_j = (j+1)\,d\,\vf^{-j}_{j+1} \ ,
    \qquad 0\leq j \leq D-2 \ , \es
   \d \,d\, \vf^{2-D}_{D-1} = 0 \ .
\ea\eqn{diff-field-delta-action}
It is easily verified then that, on the functional space $\VV$, the
operators $b$ and $\d$ obey
\eq
       d=-\lc b,\d \rc \,, \qquad \lc d,\d \rc = 0 \ ,
\eqn{delta-decomposition}
\ie $\d$ allows to decompose the exterior derivative as a BRS commutator.

Equations \equ{field-delta-action},
\equ{diff-field-delta-action} show that the operator $\d$
increases the form-degree by one unit and decreases the ghost
number by the same amount, so that it has total degree zero.
In particular from eq.\equ{delta-decomposition} it follows that
\eq
\dT = b + d = e^{\d}\, b \, e^{-\d} \ .
\eqn{delta-transform-of-b}

\section{The geometrical meaning of the zero curvature condition}

In the previous section the BRS transformations of the component fields
$\vf^{1-j}_j$ have been obtained as a consequence of the
zero curvature condition \equ{zero-curvature-cond}.

Conversely, it is very simple to show that, assuming the BRS transformations
\equ{brs-transform-cAladder} to
hold, the zero curvature condition can be derived as a consequence of the
existence of the operator $\d$. Indeed, applying $e^{\d}$ to the BRS
transformation of the ghost field $c$, \ie
\eq
 e^{\d}be^{-\d} e^{\d}c = i \, e^{\d} \, c^2,
\eqn{delta-transform-of-c}
and making use of equations \equ{A-tilde2} and
\equ{delta-transform-of-b} one gets the zero curvature condition
\eq
\dT {\wti \AA} = i  {\wti \AA}^2 .
\eqn{zero-curvature-cond-2}
This is not surprising since, as it is well known, the
ghost field $c$ identifies the so called Maurer-Cartan form of
the gauge group $G$ and its BRS transformation is nothing but
the corresponding Maurer-Cartan equation~\cite{e-h} which
is in fact a zero curvature condition. This is the geometrical meaning
of the equation \equ{zero-curvature-cond}.

\section{Cohomology of the BRS operator}

Even if the cohomology of the BRS operator in the case of a complete
ladder field has  already been studied~\cite{topol2}, let us present
here a simple derivation which may be useful for the reader.

In order to compute ththe functional space $\VV$ we introduce the filtering
operator~\cite{book,spectr} $\NN$ defined as
\eq\ba{l}
       \NN \vf^{1-j}_j = \vf^{1-j}_j  \,,\qquad 0\leq j \leq D\,, \es
       \NN d \vf^{1-j}_j = d \vf^{1-j}_j \,,
\ea\eqn{filter}
according to which the BRS operator $b$ decomposes as
\eq
b=b_0 + b_1 \ ,
\eqn{b0b1}
with
\eq\ba{l}
       b_0\,c = 0  \ , \es
       b_0\,\vf^{1-m}_m=-d\,\vf^{2-m}_{m-1} \ ,\qquad
       b_0\,d\,\vf^{2-m}_{m-1} = 0 \ ,  \qquad 1\leq m \leq D \,,
\ea\eqn{b_0-cohomology}
and
\eq
b_0^2=0 \ .
\eqn{b0nilpotent}
The usefulness of the above decomposition relies on a very
general theorem on the BRS cohomology~\cite{spectr}. The latter states that
the cohomology of the operator $b$ is isomorphic to a subspace
of the cohomology of $b_0$. We focus then on the study of the
cohomology of $b_0$.

In particular, equation \equ{b_0-cohomology} shows that all the
fields ($\vf^{1-m}_m,\,\,1\leq m \leq D$) with form degree greater
than zero and their differentials
are grouped in BRS-doublets~\cite{book,spectr,bdk}. It is known that the
cohomology does not depend on such variables.
Therefore the cohomology classes of $b_0$ depend only on the
ghost field $c$ undifferentiated, \ie they are given by elements
of the type
\eq
\o_{i_1.....i_n}c^{i_1}.....c^{i_n}
\eqn{cohomology-classes}
with $\o_{i_1.....i_n}$ arbitrary coefficients.
Moreover, from the previous theorem it follows that the
cohomology of $b$ is also given by elements of the form
\equ{cohomology-classes}  with, in addition, the restriction
that the coefficients $\o_{i_1.....i_n}$ are invariant tensors of
the gauge group~\cite{russian1,bdk,dvhtv,bbh}.

In summary, the cohomology of the BRS operator $b$ in the
complete ladder case is spanned by invariant polynomials in the
ghost field $c$ built up with monomials of the t\eq
\tr \LP \frac{c^{2n+1}}{(2n+1)!} \RP \ , \qquad n\geq 1 \ .
\eqn{b-cohomology}

\section{Solution of the descent equations}

Having computed the cohomology of the BRS operator $b$ let us
face now the problem of solving the descent equations
\eq\ba{l}
       b\,\o^{G+j}_{D-j} + d\,\o^{G+j+1}_{D-j-1} = 0 \ ,
       \qquad 0\leq j \leq D-1 \ ,   \es
       b\o^{G+D}_0 = 0 \ .
\ea\eqn{descent-eq2}
Introducing the generalized cocycle of total degree $(G+D)$
\eq
\ott^{G+D}=\sum_{j=0}^D\,\o_j^{G+D-j},
\eqn{omega-ladder}
the descent equations \equ{descent-eq2} can be cast in the more
compact form
\eq
\dT \,\ott^{G+D} = 0 \ ,
\eqn{ed-ladd}
$\dT$ being the nilpotent generalized differential of
eq.\equ{delta-transform-of-b}. Taking into account the zero
curvature condition
\eq
\dT {\wti \AA}  = i  {\wti \AA}^2
\eqn{zero-curv-cond3}
and the previous result \equ{b-cohomology} on the cohomology of the
BRS operator $b$, it follows  that the generalized monomials of the type
\eq
\tr  \frac{{\wti \AA}^{2n+1}}{(2n+1)!}   \ , \qquad n\geq 1 \ ,
\eqn{b-A-cohomology}
belongs to the cohomology of $\dT$
\eq
\dT \lp  \tr \frac{{\wti \AA}^{2n+1}}{(2n+1)!} \rp = 0 \,,\qquad
\tr \frac{{\wti \AA}^{2n+1}}{(2n+1)!} \neq \dT\, {\wti \QQ}^{2n} \ ,
\eqn{b-A-cocycle}
for any local polynomial ${\wti \QQ}^{2n}$.

It is apparent thus that a solution of the descent equations
\equ{descent-eq2} is simply provided by
\eq
\ott^{G+D} = \tr \frac{{\wti \AA}^{G+D}}{(G+D)!} \ ,
\eqn{monomial-cohomology}
which, of course, is nonvanishing only if its total degree $(G+D)$ is odd.
In fact, developping $(\tr {\wti \AA}^{G+D})$ according to the
form-degree and to the ghost-number
\eq
\lp \tr \frac{{\wti \AA}^{G+D}}{(G+D)!} \rp
\,\,\,=\,\,\,\sum_{j=0}^D\o^{G+D-j}_j\,,
\eqn{develop-monom-cohomology}
and recalling that $ \dT \, \ott^{G+D}=0 $, it is easily
verified that the $\o$'s in eq.\equ{develop-monom-cohomology}
obey to
\eq\ba{l}
       b\,\o^{G+j}_{D-j} + d\,\o^{G+j+1}_{D-j-1}=0  \ ,\es
       b\o^{G+D}_0 = 0 \ ,
\ea\eqn{descent-eq5}
\ie they solve the descent equations.

In addition, from $(\tr {\wti \AA}^{2n+1} \neq \dT\, {\wti \QQ}^{2n})$, it
follows that they provide  a nontrivial solution
\eq\ba{l}
       \o^{G+D-j}_j \neq  b\,\QQ^{G+D-1-j}_j + d\,\QQ^{G+D-j}_{j-1}\,,
       \qquad 1\leq j \leq D \,,  \es
       \o^{G+D}_0 \neq b\,\QQ^{G+D-1}_0 \ .
\ea\eqn{descent-eq6}

\noindent
In particular, for the zero form $\o^{G+D}_0$ we obtain
\eq
\o^{G+D}_0 = \tr \frac{c^{G+D}}{(G+D)!} \ .
\eqn{zero-degree-solution}
Let us remark, finally, that as a consequence of the fact that the
generalized ladder ${\wti \AA}$ is the $\d$-transform of the ghost
field $c$, ${\wti \AA} = e^{\d}\,c$, the generalized
cocycle \equ{b-A-cohomology} is the $\d$-transform of
the corresponding ghost cocycle \equ{zero-degree-solution}, \ie
\eq
\lp \tr \frac{{\wti \AA}^{2n+1}}{(2n+1)!} \rp = e^{\d} \, \tr \lp
\frac{c^{2n+1}}{(2n+1)!} \rp \ .
\eqn{b-tilde-cohomology}

\subsection{Example I: the Chern-Simons theory}

For a better understanding of the previous construction let us
discuss in details the case of the three dimensional
Chern-Simons theory, corresponding to $G=0$ and $D=3$.
This example will give us the possibility of clarifying the
meaning of the negative ghost number components
$\lp \vf^{1-j}_j \ , \,\,2\leq j \leq D \rp $
of the gauge ladder ${\wti \AA} $.
As we shall see, these fields turn out to be the
so-called external BRS sources\footnote{In the framework of
Batalin-Vilkovsky~\cite{vilk} these fields are usually called antifields.}
needed in order to properly define~\cite{book} the nonlinear transformations
of the gauge connection $A$ and of tThe external sources are then naturally
included
in the zero curvature formalism.

In a three dimensional space-time the complete gauge ladder
${\wti \AA}$ of eq.\equ{ladder-cA} takes the following form
\eq
{\wti \AA} = c + A + \g + \tau \ ,
\eqn{chern-simons-ladder}
$\g $ and $\tau $ identifying respectively the negative ghost number components
$\vf^{-1}_2$ and $\vf^{-2}_3$. From the zero curvature condition
\equ{zero-curvature-cond} one
gets the BRS transformations:
\eq\ba{l}
       b\,c = i c^2\,,  \es
       b\,A = - dc + i\lc c, A \rc \ ,\es
       b\,\g = - F + i\lc c, \g \rc \ , \es
       b\,\tau = - d\g + i\lc c, \tau \rc +i \lc A , \g \rc \ ,
\ea\eqn{brs-transf-chern-simons}
$F$ being the two-form gauge field strength $F=dA-iA^2$.
As explained before, in order to find a solution of the descent
equations
\eq\ba{l}
       b\,\o^j_{3-j} + d\,\o^{j+1}_{2-j}=0 \ ,
       \qquad 0\leq j \leq 2 \ ,  \es
       b\o^3_0 = 0 \ ,
\ea\eqn{descent-chern-simons}
it is sufficient to expand  the generalized cocycle of total
degree three
\eq
\ott^3 = \frac{1}{3!}\tr {\wti \AA}^3.
\eqn{A-monom}
After an easy computation we get
\eq
\frac{1}{3!}\tr {\wti \AA}^3 = \o^0_3 + \o^1_2 + \o^2_1 + \o^3_0 \ ,
\eqn{ttr}
with
\eq\ba{l}
       \o^3_0 = \dfrac{1}{3!}\tr c^3 \ ,\\[3mm]
       \o^2_1 = \dfrac{1}{2}\tr c^2A \ ,\\[3mm]
       \o^1_2 = \dfrac{1}{2}\tr \lp c^2\g + cA^2 \rp \ ,\\[3mm]
       \o^0_3 = \dfrac{1}{2}\tr \lp c^2\tau + cA\g + c\g A +
                \dfrac{A^3}{3} \rp \ .
\ea\eqn{ldr}
{}From
\eq
 -i \tr \lp c^2\tau + cA\g + c\g A \rp = - \tr AF + b\,\tr \lp c\tau +
A\g \rp + d\,\tr c\g  \ ,
\eqn{trtrt}
the three-form $\o^0_3$ can be rewritten as
\eq
\o^0_3 = \frac{-i}{2}\tr (AF + i\frac{A^3}{3}) +
\frac{i}{2}b\,\tr (c\tau + A\g) + \frac{i}{2}d\,\tr c\g  \ ,
\eqn{actinv}
yielding thus the invariant action
\eq
S = i\int \o^0_3 = \frac{1}{2}\int \tr (AF + i\frac{A^3}{3}) -
\frac{1}{2}b\, \int \tr (c\tau + A\g)  \ ,
\eqn{chern-simons-action}
which is easily recognized to be the so-called truncated action~\cite{book}
of the fully quantized Chern-Simons gauge theory. In
particular, one sees that the components $(\g ,\tau )$ of the gauge
ladder \equ{chern-simons-ladder} are the  BRS external
sources corresponding to the nonlinear transformations of the
fields $A$ and $c$.
\section{Coupling with matter fields}

The zero curvature formalism can be extended to include the case
in which the gauge fields are coupled to matter fields whose
quantization requires the introduction of a complete ladder
matter multiplet.
A typical example of this kind of coupling is given by the
topological $BF$ systems~\cite{topol-rep,hhs} whose classical action reads
\eq
\tr \int_{\MM^D}\BB^0_{D-2}F \ ,
\eqn{bf-action}
where $F$ is the two-form gauge curvature,
$\BB^0_{D-2}$ is a $(D-2)$ form with ghost number zero and
$\MM^D$ a $D$-dimensional manifold without boundaries.

In the next section we shall discuss another example of matter
system, namely the $B$-$C$ ghost system of the string theory,
whose action is not directly given in terms of differential forms.
Nevertheless we shall see that this model, although different
from the $BF$ systems, actually shares many properties of the
latters.

The inclusion of the matter fields goes as follows
(see also ref.~\cite{ncab}): we introduce
a set of fields $(\BB^{D-2}_0, \BB^{D-3}_1,...,
\BB^1_{D-3}, \BB^{-1}_{D-1}, \BB^{-2}_D )$ which together with the
matter field $\BB^0_{D-2}$ give rise to a complete ladder ${\wti \BB}$
of total degree $(D-2)$, \ie
\eq
\BT = \sum_{j=0}^D \BB^{D-2-j}_j \ .
\eqn{B-ladder}
The BRS transformations of the various components of this ladder areobtained by
requiring that $\BT $ is covariantly constant with
respect to the generalized
covariant derivative $\DT = \dT - i [ {\wti \AA} ,\,\,
\,\,] $,
\eq
\DT \BT = \dT \BT - i[ {\wti \AA} , \BT ] = 0 \ .
\eqn{covar-deriv-B-ladder}
This condition, when expanded in terms of the form degree and of
the ghost number, gives in fact the following nilpotent transformations:
\eq\ba{l}
    b\,\BB^{D-2}_0 = i \lc c , \BB^{D-2}_0 \rc \ , \es
    b\,\BB^{D-2-j}_j = -d\,\BB^{D-1-j}_{j-1} + i\dsum{m=0}{j}
    \lc \,\vf^{1-m}_m\,,\,\BB^{D-2-j+m}_{j-m} \rc \,, \qquad
    1\leq j \leq D \ .
\ea\eqn{brs-transform-B-ladder}
Repeating the same procedure of Sect.4 and making use of the general
results of refs.~\cite{russian1,bdk,dvhtv,bbh}, one easily checks
 that with the
inclusion of the matter ladder the cohomology of the BRS
operator is  given by polynomials in the undifferentiated zero form
ghosts $(c,\BB^{D-2}_0)$ built up with factorized monomials of
the type
\eq
\lp \tr \frac{c^{2n+1}}{(2n+1)!}\rp \cdot \tr \lp \BB^{D-2}_0\rp^m
 \ , \qquad  m\geq 1 \ .
\eqn{cohom-brs-with-matter-monomials}
In much the same way as the gauge ladder ${\wti \AA}$, the operator
$\d$ extends to the matter multiplet $\BT $ by means of
\eq
\BT = e^{\d} \, \BB^{D-2}_0 \ ,
\eqn{B-ladder2}
\ie
\eq\ba{l}
    \d \, \BB^{D-2-j}_j = (j+1)\BB^{D-3-j}_{j+1} \ ,
      \qquad 0\leq j \leq D-1 \,, \es
   \d \BB^{-2}_D = 0 \ ,
\ea\eqn{matter-field-delta-action}
and
\eq\ba{l}
    \d \,d\, \BB^{D-2-j}_j = (j+1)\,d\,\BB^{D-3-j}_{j+1} \ ,
    \qquad 0\leq j \leq D-2 \,, \es
   \d \,d\, \BB^{-1}_{D-1} = 0 \ ,
\ea\eqn{diff-matter-field-delta-action}
so that the algebraic relations
\eq
              d=-\lc b,\d \rc\,,\qquad  \lc \d , d \rc = 0 \ ,
\eqn{dddelta}
are fulfilled.

For what concerns the cohomology of the generalized operator
$\dT $ of eqs.\equ{zero-curv-cond3} and \equ{covar-deriv-B-ladder}, it is
immediately seen
from eq.\equ{cohom-brs-with-matter-monomials} that it is spanned
by factorized monomials in the ladders ${\wti \AA}$ and $\BT $ of the type
\eq
\lp \tr \frac{{\wti \AA}^{2n+1}}{(2n+1)!}\rp \cdot  \lp \tr  \BT^m  \rp \ .
\eqn{cohom-brs-with-matter-monomials2}
As already discussed in the previous section, the expansion of
the above expression \equ{cohom-brs-with-matter-monomials2} in
terms of the form degree and of the ghost number yields a solution of
the descent equations \equ{descent-eq2} in the presence of a
 matter field ladder, reproducing thus the results already
established in~\cite{topol2}.
Again
\eq
\lp \tr \frac{{\wti \AA}^{2n+1}}{(2n+1)!}\rp \cdot \tr  \BT^m
= e^{\d}\lp \tr \frac{c^{2n+1}}{(2n+1)!}\rp \cdot \lp \tr \lp \BB^{D-2}_0
\rp^m \rp
\eqn{matter-solution-cohomology}
which shows that the cohomology of $\dT$ is the $\d$-transform
of that of the BRS operator $b$.
Let us conclude this section by remarking that in a space-time
of dimension $(D\geq 2)$ the gauge ladder ${\wti \AA} $ contains
$(D-1)$ components of negative ghost number, \ie
$(\vf^{-1}_2,...,\vf^{1-D}_D)$, while the matter ladder $\BT$
contains $(D-2)$ components with positive ghost number, \ie $\lp
\BB^{D-2}_0, \BB^{D-3}_1,..., \BB^1_{D-3}\rp
$, and two components of negative ghost number, namely
$\lp \BB^{-1}_{D-1}, \BB^{-2}_D\rp $.

\noindent
These fields turn out to possess the following meaning. The set $\lp
\BB^{D-2}_0,
\BB^{D-3}_1,..., \BB^1_{D-3}\rp$
identifies the well known tower of ghosts for ghosts needed for
the quantization of the $BF$ systems. The components
$\lp \vf^{-2}_3,...,\vf^{1-D}_D\rp $ are then the corresponding
$(D-2)$ external sources (or antifields) associated to the
nonlinear transformations of the ghosts for ghosts (see
eq.\equ{brs-transform-B-ladder}), while $\vf^{-1}_2$ is the
external source for the $(D-2)$ form $\BB^0_{D-2}$.
Finally $\lp \BB^{-1}_{D-1}, \BB^{-2}_D \rp $ are the
sources corresponding to the first two components of the gauge
ladder, i.e. $c$ and $A$. We thus see that in the case of the
$BF$ systems the external sources are exchanged~\cite{topol2}, i.e. the
sources for the quantized components of the matter ladder are
grouped into the gauge ladder and vice versa.

Let us also recall, for completeness, that the truncated action
(including the ghosts for ghosts and the external sources) for the
$BF$ systems can be cast in the simple form~\cite{ncab}
\eq\ba{l}
S= \left.
   {\tr \dint_{\MM^D}\BT \lp d {\wti \AA} - i {\wti \AA}^2 \rp}
   \right|^0_D \,
 = - \left.
   {\tr \dint_{\MM^D}\BT \, b \, {\wti \AA}}
   \right|^0_D  \ ,
\ea\eqn{BF-action}
where $|^0_D$ means the restriction to terms of ghost number $0$
and form degree $D$. The equality in eq.\equ{BF-action}
stems from the zero
curvature condition \equ{zero-curvature-cond}.

In particular, using equation \equ{covar-deriv-B-ladder},
expression \equ{BF-action} is easily proven to be invariant
under the action of the operator $b$,
\eq
b\,S=0 \ ,
\eqn{Sbaction}
this equation expressing
the content of the Slavnov-Taylor (or Master Equation) identity.
\section{Example II: The $B$-$C$ ghost system}

We present here, as another interesting example of matter
system, the zero curvature formulation of the two dimensional $B$-$C$ model
whose action reads
\eq
S_{B-C}=\int dzd{\bar z}\, B\,\bpart C \ ,
\eqn{bc-action}
where the fields $B=B_{z\,z}$ and $C=C^z$ are anticommuting and
carry respectively ghost number $-1$ and $+1$.

It should be noted that, unlike the previous examples, the fields
appearing in the action \equ{bc-action} are not naturally associated to
differential forms. However we shall see that, in spite of the fact
that these fields do not gsystem turns out to possess the same algebraic
features of the $BF$ models.
The action \equ{bc-action} is recognized to be the ghost part of
the quantized bosonic string action~\footnote{Expression \equ{bc-action} is
usually accompanied by its complex conjugate. However, the inclusion of the
latter in the present framework does not require any additional difficulty.}
which, as it is well know, is
left invariant by the following nonlinear BRS transformations
\eq\ba{l}
s \,C = C\,\partial \,C  \ , \es
s \,B = - \lp \partial \, B \rp C - 2\,B\,\partial \, C \ .
\ea\eqn{brs-bc-transf}
In particular, the right hand-side of the BRS transformation of the
field $B$ is recognized to be the component $T_{z\,z}$ of the
energy-momentum tensor corresponding to the action \equ{bc-action}, this
property allowing for a topological interpretation of the model.

Transformations \equ{brs-bc-transf} being nonlinear, one needs
to introduce two external invariant sources $\m =
\m^z_{\,\,\,{\bar z}}$ and $L = L_{z\,z\,{\bar z}}$ of ghost
number respectively $0$ and $-2$
\eq
S_{{\it ext}} = \int dzd{\bar z} \lp \m \,s\,B  + L\,s\,C \rp \ .
\eqn{bc-ext-action}
The complete action
\eq
S=S_{B-C} + S_{{\it ext}}
\eqn{total-bc-action}
obeys thus the classical Slavnov-Taylor identity
\eq
\int dzd{\bar z} \lp \dfrac{\d S}{\d B}\dfrac{\d S}{\d \m} +
  \dfrac{\d S}{\d L}\dfrac{\d S}{\d C} \rp = 0 =
\frac{1}{2}\,b\,S \ ,
\eqn{bc-Slav-Taylor-ident}
$b$ denoting the nilpotent linearized operator
\eq\ba{l}
 b \, = \, \dint dzd{\bar z} \lp \dfrac{\d S}{\d B}\dfrac{\d }{\d \m} +
 \dfrac{\d S}{\d \m}\dfrac{\d }{\d B} + \dfrac{\d S}{\d L}\dfrac{\d }{\d C}
+ \dfrac{\d S}{\d C}\dfrac{\d }{\d L} \rp  \ .
\ea\eqn{b-operator-def}
The operator $b$ acts on the fields and on the external sources
in the following way
\eq\ba{l}
b\,C = s\,C = C\,\partial \,C  \ , \es
b\,\m = \bpart C +  \lp \partial \, \m \rp \, C
- \m \, \lp \partial \, C \rp \ ,
\ea\eqn{cm-brs-transf}
and
\eq\ba{l}
b\,B = s \,B = - \lp \partial \, B \rp C  - 2\,B\,\partial \, C \ , \es
b\,L = \bpart \, B - \lp 2\,B \rp \partial \m  - \m \partial\, B +
\lp \partial \, L \rp C + 2\,L\,\partial \,C \ .
\ea\eqn{BL-brs-transf}
It should be noted that, due to the fact that the BRS
transformation of $B$ is the component $T_{z\,z}$ of the
energy-momentum tensor, the differentiation with respect to the
external source $\m$ of the Legendre transformation of the complete action
\equ{total-bc-action}
\eq
Z(j,\m , L) = S + \int dzd{\bar z}\lp j_C\,C + j_B\,B \rp \ ,
\eqn{Legendre-transf-funct}
allows to obtain
the Green functions with insertion of $T_{z\,z}$. In other words,
the Slavnov-Taylor identity \equ{bc-Slav-Taylor-ident} is the
starting point for the algebraic characterization of the
energy-momentum current algebra.

\noindent
Introducing now the two functional operators~\cite{corde}
\eq
\WW = \dint dzd{\bar z} \dfrac{\d }{\d C} \ , \qquad
{\overline \WW} = \dint dzd{\bar z} \lp  \m \,\dfrac{\d }{\d C} +
L\,\dfrac{\d }{\d B}\rp \ ,
\eqn{slavnov-tay-operators}
one easily proves that
\eq
\d = dz\,\WW + d{\bar z}\,{\overline \WW}
\eqn{bc-action2}
obeys to
\eq
       d=-\lc b,\d \rc \ , \qquad \lc d,\d \rc=0 \ ,
\eqn{delta-decomposition4}
$d$ being the exterior derivative $d= dz\partial + d{\bar z}{\bpart}$.
We have thus realized the decomposition
\equ{delta-decomposition}. In order to derive the transformations
\equ{cm-brs-transf}, \equ{BL-brs-transf} from a zero curvature
condition we proceed as before and we define the analogue of the
gauge ladder \equ{ladder-cA} as
\eq
{\tilde C}^z = e^{\d}\,C^z = C^z +  dz + d{\bar z}\m^z_{\bar z}\ .
\eqn{cc-ladder}
Introducing then the holomorphic generalized vector field
${\tilde C}={\tilde C}^z\,\partial_z$, it is easily checked that
equations \equ{cm-brs-transf} can be cast in the form of a zero
curvature condition
\eq
\dT \,{\tilde C} = \frac{1}{2}\lc {\tilde C}, {\tilde C} \rc \, =
      \LL_{{\tilde C}}\,{\tilde C}  \ ,
\eqn{bc-zero-curvature}
where, as usual, $\dT$ is the operator
\eq
\dT = e^{\d}\,b\,e^{-\d}=b+d \,,
\eqn{d-tilde-def}
and $\LL_{{\tilde C}}$ denotes the Lie derivative with respect to
the vector field\footnote{Of course, the bracket
$[ {\tilde C}, {\tilde C}]$ in eq.\equ{d-tilde-def} refers
now to the Lie bracket of vector fields.} ${\tilde C}$.

Concerning now the second set of transformations
\equ{BL-brs-transf}, we define the matter ladder $\BT_{z\,z}$ as
\eq
\BT_{z\,z} = e^{\d}\BB_{z\,z}
=\BB_{z\,z} + d{\bar z}L_{z\,z\,{\bar z}} \ .
\eqn{bc-matter-ladder}
To expression \equ{bc-matter-ladder} one can naturally associate
the generalized holomorphic quadratic differential
\eq
\BT =  \BT_{z\,z}dz\otimes dz \ .
\eqn{quadratic-diff-B-ladder}
Therefore, transformations \equ{BL-brs-transf} can be rewritten
as
\eq
\dT \,\BT - \LL_{{\tilde C}}\,\BT = 0 \ .
\eqn{B-ladder-curv-zero}
This equation is the analogue of the covariantly
constant matter condition \equ{covar-deriv-B-ladder} and together with
the equation \equ{bc-zero-curvature} completely characterize the
$B$-$C$ system.
One has to remark that, as it happens in the case of the $BF$
models, the external sources $(\m , L)$ are interchanged, \ie
the source $\m $ associated to the nonlinear transformation of
$B$ belongs to the gauge ladder ${\tilde C}$ and vice versa.

Let us consider now the problem of identifying the anomalies which affect the
Slavnov-Taylor identity \equ{bc-Slav-Taylor-ident}
at the quantum level. We look then at the solution of the
descent equations
\eq\ba{l}
b\,\o^1_2 + d\,\o^2_1 = 0 \ ,\es
b\,\o^2_1 + d\,\o^3_0 = 0 \ ,\es
b\,\o^3_0  = 0 \ .
\ea\eqn{bc-desc-eq}
As it has been proven in refs.~\cite{corde,cordeb}, the cohomology  of the BRS
operator in the sector of the zero-forms with ghost number three
contains, in the present case,  a unique element given by
\eq
\o^3_0 = C\,\partial \,C\,\partial^2 \,C \ .
\eqn{zero-level-bc-cohom}
{}From the zero curvature condition
\equ{bc-zero-curvature}, it follows then that the generalized cocycle of total
degree three
\eq
{\tilde \o}^3 = {\tilde C}\,\partial \,{\tilde C}\,\partial^2 \,{\tilde C} \ ,
\eqn{ladder-cocycle-bc}
belongs to the cohomology of $\dT $.
The expansion of ${\tilde \o}^3$ will give thus a solution of
the ladder \equ{bc-desc-eq}, \ie
\eq
{\tilde \o}^3 = \o^3_0 + \o^2_1 + \o^1_2 \ ,
\eqn{bc-ladder-solution}
with $\o^2_1$, $\o^1_2$ given respectively by
\eq\ba{l}
\o^2_1 = \lp C\,\partial \,C \partial^2 \,\m -
C \, \partial^2 \,C\,\partial \m +
\m \,\partial \,C \partial^2 \,C \rp d{\bar z} +
\lp \partial \,C \rp \lp \partial^2 \,C \rp dz  \ , \es
\o^1_2 = \lp - \partial \,C \partial^2 \,\m +
\partial \,\m  \partial^2 \,C \rp dz\wedge d{\bar z} \ .
\ea\eqn{explicit-bc-ladder-solut}
In particular,
\eq
\int \o^1_2 = 2\,\int dzd{\bar z}\,C \partial^3 \m
\eqn{bc-anomaly}
is recognized to be the well known two-dimensional
diffeomorphism anomaly characterizing the central charge of the
energy-momentum current algebra.

Let us conclude by remarking that the complete $B$-$C$
action \equ{total-bc-action} can be written, in perfect analogy
with eq. \equ{BF-action}, as
\eq\ba{l}
S = \left. {\dint \BT_{z\,z} \lp d {\tilde C}^z - {\tilde C}^z
\partial {\tilde C}^z \rp  dz}
 \right|^0_2  \,  = \,
-  \left.  {\dint \BT_{z\,z} \, b \, {\tilde C}^z dz}
 \right|^0_2 \ ,
\ea\eqn{BF-action5}
showing that the $B$-$C$ model can be interpreted as a kind of
two-dimensional $BF$ system.
\section*{Conclusion}

The zero curvature formulation of models characterized by means of a complete
ladder field can be obtained as a consequence of the existence of the
operator $\d$ realizing the decomposition \equ{delt-def}.
Moreover, the zero curvature condition enables us to encode into a
unique equation all the relevant informations concerning the BRS
cohomology classes.


\end{document}